\documentclass[twocolumn]{aastex631}

\newcommand\src{1RXS~J170849.0-400910}
\newcommand{\srcshort}{1RXS~J1708}
\newcommand{\ixpe}{{\it IXPE}}
\newcommand{\srcold}{4U 0142+61}
\newcommand{\ixpeobssim}{\textsc{ixpeobssim}}

\begin{document}
\AuthorCollaborationLimit=400

\title{A strong X-ray polarization signal from the magnetar \src }

\correspondingauthor{Silvia Zane}
\email{s.zane@ucl.ac.uk}

\author[0000-0001-5326-880X]
{Silvia Zane}
\affiliation{Mullard Space Science Laboratory, University College London, Holmbury St Mary, Dorking, Surrey RH5 6NT, UK}

\author[0000-0002-1768-618X]{Roberto Taverna}
\affiliation{Dipartimento di Fisica e Astronomia, Universit\`{a} degli Studi di Padova, Via Marzolo 8, I-35131 Padova, Italy}

\author[0000-0001-5848-0180]{Denis Gonz\'alez--Caniulef}
\affiliation{Department of Physics and Astronomy, University of British Columbia, Vancouver, BC V6T 1Z1, Canada} 
\affiliation{CITA National Fellow} 
\affiliation{Institut de Recherche en Astrophysique et Plan\'etologie
9 avenue du Colonel Roche
BP 44346
31028 Toulouse CEDEX 4
France}

\author[0000-0003-3331-3794]{Fabio Muleri}
\affiliation{INAF Istituto di Astrofisica e Planetologia Spaziali, Via del Fosso del Cavaliere 100, I-00133 Roma, Italy}

\author[0000-0003-3977-8760]{Roberto Turolla}
\affiliation{Dipartimento di Fisica e Astronomia, Universit\`{a} degli Studi di Padova, Via Marzolo 8, I-35131 Padova, Italy}
\affiliation{Mullard Space Science Laboratory, University College London, Holmbury St Mary, Dorking, Surrey RH5 6NT, UK}

\author[0000-0001-9739-367X]{Jeremy Heyl}
\affiliation{Department of Physics and Astronomy, University of British Columbia, Vancouver, BC V6T 1Z1, Canada} 

\author[0000-0002-7752-9389]{Keisuke Uchiyama}
\affiliation{
RIKEN Nishina Center, 2-1 Hirosawa, Wako, Saitama 351-0198, Japan}
\affiliation{Department of Physics, Tokyo University of Science, 1-3 Kagurazaka, Shinjuku, Tokyo 162-8601, Japan}

\author[0000-0002-0940-6563]{Mason Ng}
\affiliation{MIT Kavli Institute for Astrophysics and Space Research, Massachusetts Institute of Technology, 77 Massachusetts Avenue, Cambridge, MA 02139, USA}

\author[0000-0002-8801-6263]{Toru Tamagawa}
\affiliation{RIKEN Cluster for Pioneering Research, 2-1 Hirosawa, Wako, Saitama 351-0198, Japan}
\affiliation{
RIKEN Nishina Center, 2-1 Hirosawa, Wako, Saitama 351-0198, Japan}
\affiliation{Department of Physics, Tokyo University of Science, 1-3 Kagurazaka, Shinjuku, Tokyo 162-8601, Japan}

\author[0000-0002-4770-5388]{Ilaria Caiazzo}
\affiliation{ Cahill Astrophysics, 1216 California Blvd,
MC 350-17, Caltech, Pasadena, CA 91125}

% ...



\begin{abstract} Magnetars are the most strongly magnetized neutron stars, and one of the most promising targets for X-ray polarimetric measurements. We present here the first Imaging X-ray Polarimetry Explorer (\ixpe) observation of the magnetar \src, jointly analysed with a new Swift observation and archival NICER data. The total (energy and phase integrated) emission in the 2--8~keV energy range is linerarly polarized, at a $\sim 35$\% level. The phase-averaged polarization signal shows a marked increase with energy, ranging from $\sim$ 20\% at 2--3~keV up to $\sim$ 80\% at 6--8~keV, while the polarization angle remain constant. This indicates that radiation is mostly polarized in a single direction. The spectrum is well reproduced by a combination of either two thermal (blackbody) components or a blackbody and a power law. Both the polarization degree and angle also show a variation with the spin phase, and the former is almost anti-correlated with the source counts in the 2--8~keV and 2--4~keV bands. 
We discuss the possible implications and interpretations, based on a joint analysis of the spectral, polarization and pulsation properties of the source.
A scenario in which the surface temperature is not homogeneous, with a hotter cap covered by a gaseous atmosphere and a warmer region in  a condensed state, provides a 
satisfactory description of  both the phase- and energy-dependent spectro-polarimetric data. 
The (comparatively) small size of the two emitting regions, required to explain the observed pulsations, does not allow to reach a robust conclusion about the presence of vacuum birefringence effects.


\end{abstract}

\keywords{X-rays: stars --- stars: magnetars ---
techniques: polarimetric}

\section{Introduction}
\label{sec:int}

The launch of the NASA-ASI {\it Imaging X-ray Polarimetry Explorer} \cite[\ixpe;][]{Weisskopf2022} in December 2021 opened a new window to our view of the X-ray sky by adding polarimetry to spectroscopy, timing and imaging as a tool to interpret astrophysical X-ray sources. \ixpe\ has been in operation for nearly one year now, and already observed more than $30$ X-ray sources belonging to different classes.

Particularly interesting for \ixpe\ observations are magnetars, a class of isolated neutron stars (NSs) that are powered by their huge magnetic field \cite[][]{td92,td93}. They are characterized by an X-ray luminosity $L\approx 10^{31}$--$10^{36}$~erg~s$^{-1}$, spin period $P \approx 1$--$12 $~s, and  period derivative $\dot P \approx 10^{-13}$--$10^{-11}$~s~s$^{-1}$, implying a dipole field $B\approx 10^{14}$--$10^{15}$ G. Magnetars are extremely active sources over different ranges of luminosities and timescales, from the short-lived 
X-ray bursts and powerful giant flares to outbursts, during which their persistent X-ray flux suddenly increases by a factor of $\approx 10$--$1000$
and then gradually decays over months/years \cite[e.g.][]{rea+esp11,cotizelati+18}. 

Magnetar X-ray spectra below $10$~keV are typically well reproduced by a two-component model,
comprising either two thermal (blackbody, BB) components or a thermal component plus a non-thermal (power-law, PL) one. While the BB component(s) is believed to originate from (regions of) the cooling star surface, the PL one is usually interpreted as due to resonant cyclotron scattering (RCS) of thermal photons off  magnetospheric currents flowing in a twisted magnetic field. In addition, magnetars exhibit a powerful and highly pulsed non-thermal emission at higher energies, up to $\sim 100$--$200$~keV, first discovered by {\it INTEGRAL} \cite[see e.g.][for reviews]{turolla+15,kaspi+belo17}. 

Strongly magnetized sources, which electromagnetic emission is expected to be highly polarized, are ideal targets for X-ray polarimetry. 
Emission from highly magnetized neutron stars  is expected to be linearly polarized in two normal modes, the ordinary (O) and extraordinary (X) ones, with the polarization electric vector
either parallel or perpendicular to the plane of the photon direction and the (local) magnetic field. The degree of polarization depends on the properties of the emission region, its geometry, magnetic field strength and orientation,  and on the dominant radiative processes. This makes X-ray polarimetry  a new, powerful tool to probe the magnetospheric topology of the magnetar, 
the state of matter in the star crust, including  whether the star has a bare, condensed surface  or is surrounded by a gaseous atmosphere \cite[see e.g.][and references therein]{taverna+20,Caiazzo2022}, and even to test the properties of the QED magnetized vacuum around the star, such as its birefringence  \citep{heyl2000,heyl2002}.

\ixpe\ first observed a magnetar source, the bright Anomalous X-ray Pulsar (AXP) \srcold, in February 2022. The polarization signal was clearly detected at $\sim 13\%$ level, with both the polarization degree and angle exhibiting a strong dependence on the energy \cite[][]{taverna+22}. In this paper we report on the \ixpe\ observation of a second magnetar, the AXP \src.

\section{Source properties and observations} \label{obs}

\src\  (\srcshort\ hereafter) was first identified in the {\it ROSAT} All Sky Survey \cite[][]{voges+99}, but only a few years later $\sim 11$~s pulsations were discovered by {\it ASCA} \cite[][]{sugizaki+97}. Both the large period derivative ($\dot P \sim 2 \times 10^{-11}$ s~s$^{-1}$) and the upper limits inferred for a possible optical counterpart confirmed the source to be a  member of the magnetar class \cite[][]{israel+99}. The value of the period derivative, if interpreted as due to a rotating dipole in vacuum, yields a dipole magnetic field strength of $\sim 4-5 \times 10^{14}$~G. The source presents an erratic timing behavior, characterized by frequent glitching activity that interrupt periods of steady spin-down \cite[][]{DibKaspi2014}. The X-ray spectrum has been fitted with an absorbed BB+PL model, with (slightly variable) photon index $\Gamma \sim  2.6$ and blackbody temperature $kT\sim 0.45$~keV \cite[][]{rea+03,rea+07,Krawczynski+22}. 
The source is radio-quiet. Hard, strongly pulsed X-ray emission extending up to $\sim$ 150~keV has been detected from \srcshort\ with {\it INTEGRAL} \cite[][]{kuiper+1996}; 
 a marked variability in the hardness ratio is also observed \citep{Gotz+07}.

The distance of the source is uncertain. \srcshort\ lies in the Galactic plane, and the large column density inferred from X-ray data, $n_\mathrm H\sim 1.36\times 10^{22}\ \rm{cm}^{-2}$,  suggests a distance in the $5$--$10$ kpc range \cite[][]{israel+99,rea+07}, although a smaller value cannot be ruled out, as hinted from spectral fits with atmospheric models or from reddening measures based on 2MASS stars \cite[][]{perna+2001, durant+06}.
In this paper we present the results from the 
\ixpe\  observation of  \srcshort\, performed from 2022-09-19 05:08 UTC to 2022-09-29 12:00 UTC and from 2022-09-30 12:52:02 UTC to 2022-10-08 11:17:40 UTC. We  complemented the analysis with data from a Swift-XRT \citep{SwiftXRT}  target of opportunity observation,  performed on  2022-10-20  for a total of $1020\,\mathrm s$, and with the most recent NICER \citep{nicer} data publicly available (see Appendix~\ref{appA} for details). Our observations and the methods used for data processing are described in Appendix~\ref{appA}.

\section{Phase-integrated polarimetric and spectropolarimetric analysis} \label{polspec}

\begin{figure*}[th!]
\begin{center}
\includegraphics[width=0.8\textwidth]{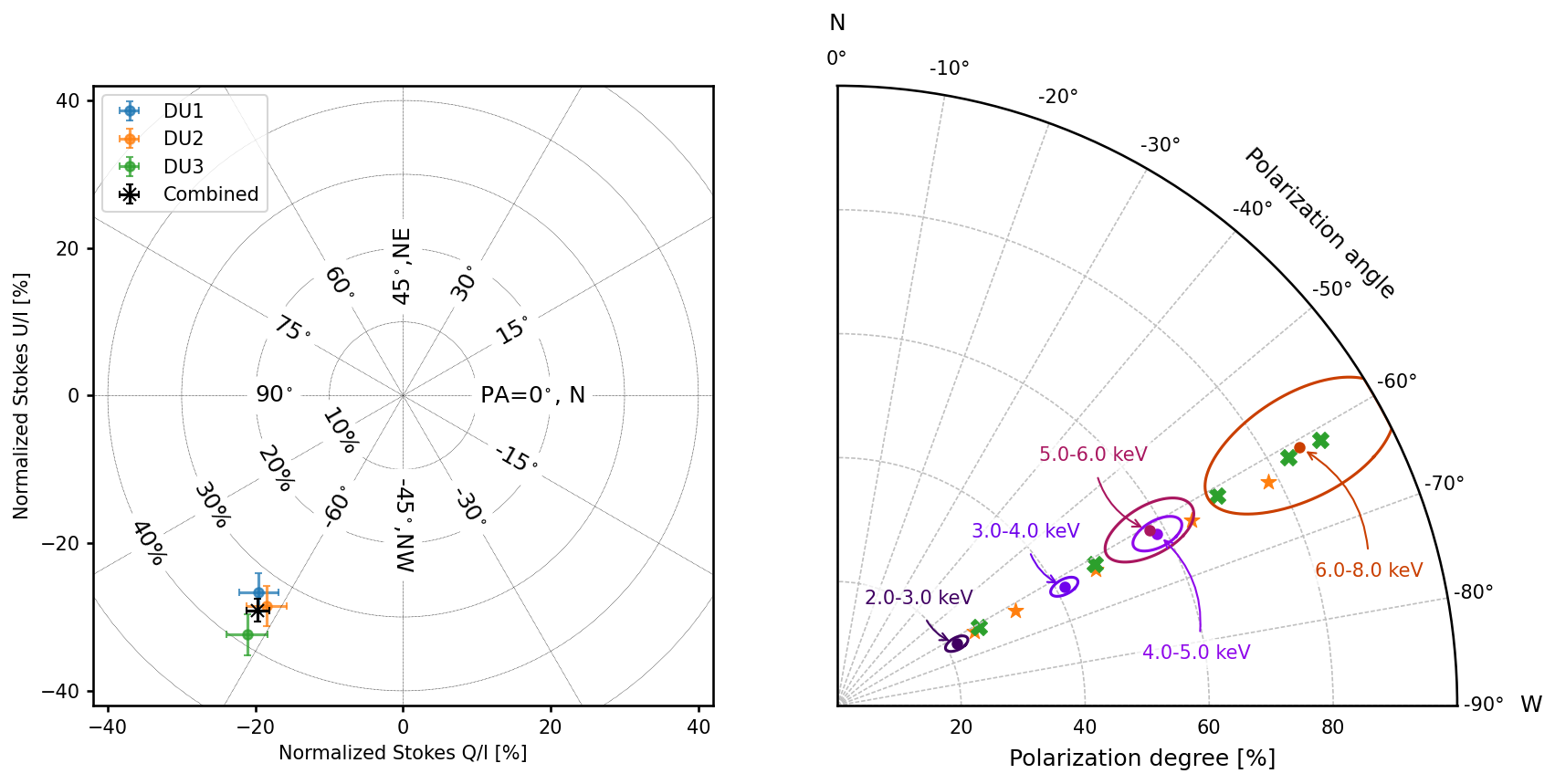} 
\end{center}
\caption{Left: Phase- and energy-averaged linear 
polarization of \srcshort, in terms of the 
normalized Stokes parameters. 
Uncertainties are shown at the 68.3\% ($1\, \sigma$) confidence level.
Circles with center in the origin and increasing radii correspond to increasing values of the polarization degree while different  values of the azimuth to different polarization angles. 
Right: Polarization of \srcshort{} in different energy bands, computed with \ixpeobssim. Contours identify  the 50\% confidence regions for the joint measurement of the polarization degree and angle (which are not independent variables), accounting for 
statistical fluctuations only. 
Orange stars and green crosses show, for the same energy bins,  
the prediction of the belt+cap and cap+cap models, respectively
(see \S \ref{discuss} for details) 
\label{fig:pol_vs_energy_new}}
\end{figure*}

No variations in either the source or background were found during the two \ixpe\ observations (see Appendix \ref{ixpedata}), therefore  the combined data are used in the following analysis. 
The phase integrated  and energy integrated polarization in the $2.0$--$8.0$~keV interval, which is the nominal \ixpe\ energy range, is shown in Figure
~\ref{fig:pol_vs_energy_new} (left panel) 
for each detector unit (DU), and for their combination. 
The values of the Stokes parameters were derived with the \ixpeobssim\ suite \citep[][]{Baldini2022}\footnote{\url{https://github.com/lucabaldini/ixpeobssim}}, and they are  weighted by the (energy-dependent) effective area of the instrument. This allows us to correct the measured polarization, which is convolved with the spectral response of the instrument, and to obtain the polarization of the incoming radiation. The circles and the radial lines in  Figure~\ref
{fig:pol_vs_energy_new} mark the loci of constant polarization degree, $\mathrm{ PD}=\sqrt{Q^2+U^2}/I$ and polarization angle, $\mathrm{PA}=\arctan{(U/Q)}/2$, respectively.
Since the Stokes parameters are (quasi) normally distributed, we report the one-dimensional uncertainties at 68.3\% confidence level ($1\,\sigma$). 

The measured polarization angle is $\sim$ 60$^{\circ}$ West of North, and the phase- and energy-integrated polarization degree is remarkably high, $\sim35\%$, that is more than 2.5~times the average X-ray polarization of the only other magnetar observed to date \citep{taverna+22}. 

Even higher polarization degrees are detected by binning the data in five energy intervals, as reported in the right panel of Figure~\ref{fig:pol_vs_energy_new}; here the contours show the 50\% confidence regions in the polarization degree vs. polarization angle plane (see also Table~\ref{poltable}). The polarization degree increases from $\sim20\%$ between $2$ and $3$ keV to $\sim80\%$ in the $6$--$8$~keV energy range. 
The polarization angle remains roughly  constant in all energy bins.


\begin{table*}
\begin{tabular}{l|lllll|l}
\hline
{} &            2--3 keV &            3--4 keV &            4--5 keV &             5--6 keV &                   6--8 keV &            2--8 keV \\
\hline
PD - sum [\%]  &   21.7$^{+1.7}_{-1.7}$ &   41.3$^{+2.0}_{-2.0}$ &   58.6$^{+3.7}_{-3.7}$ &   57.7$^{+6.8}_{-6.8}$ &       85$^{+15}_{-15}$ &   35.1$^{+1.6}_{-1.6}$ \\
PD S/N         &          12.9 $\sigma$ &          20.2 $\sigma$ &          15.8 $\sigma$ &           8.5 $\sigma$ &           5.8 $\sigma$ &          22.5 $\sigma$ \\
PA - sum [deg] &  $-62.6^{+2.2}_{-2.2}$ &  $-62.4^{+1.4}_{-1.4}$ &  $-61.8^{+1.8}_{-1.8}$ &  $-60.7^{+3.3}_{-3.3}$ &  $-60.8^{+4.7}_{-4.7}$ &  $-62.1^{+1.3}_{-1.3}$ \\
\hline
\end{tabular}
\caption{
Values of the measured polarization degree and angle, obtained with the
\ixpeobssim~software suite. Reported values correspond to the sum of the three DUs (the measures of the single DUs are consistent with each other within errors). Uncertainties are obtained at 68.3\%
confidence level, assuming that the polarization degree and angle are
independent. Signal-to-noise is calculated by dividing the polarization
degree obtained by combining the results of the three telescopes on-board \ixpe\  by its uncertainty.
}
\label{poltable}
\end{table*}

We performed a spectropolarimetric analysis making use of the 
\ixpe, {\it Swift}-XRT and NICER 
observations (the latter performed within a month; see Appendix~\ref{appA}). We extracted 
{\it Swift} and NICER energy spectra with the standard
data analysis tools and used \ixpeobssim{} to derive the 
\ixpe\ $I$, $Q$ and $U$ energy spectra. We use XSPEC (version: 12.13.0,  \citealt{Arnaud1996}) for a joint spectropolarimetric fit of the $I$ spectra from all three observatories and the $Q$ and $U$ spectra from 
{\it IXPE}.

\begin{figure*}[th!]
\includegraphics[width=\textwidth]{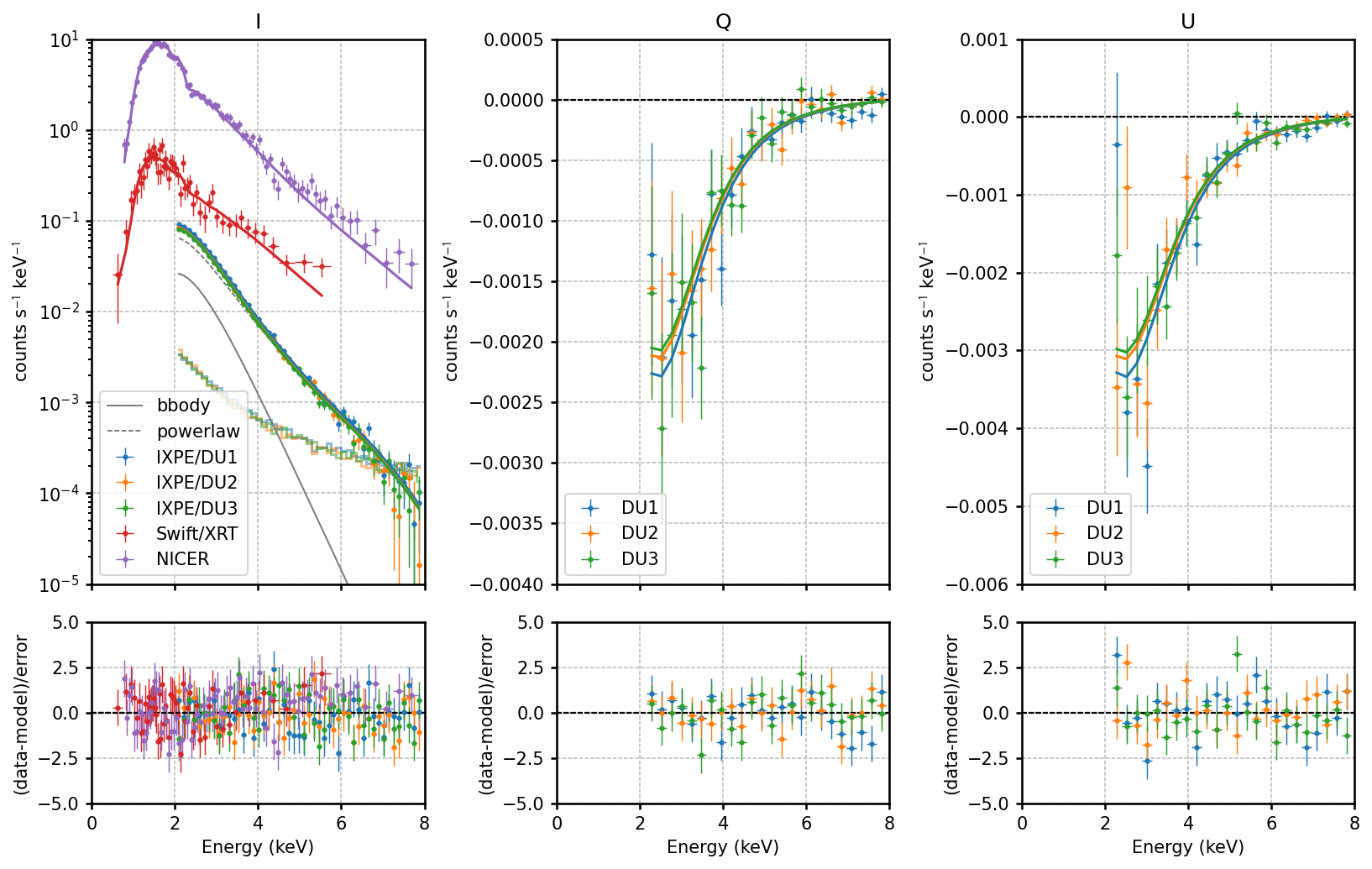} 
\caption{Joint spectropolarimetric fit of {\it IXPE}, {\it Swift}/XRT and NICER data. 
The Stokes $I$, $Q$ and $U$ energy spectra are obtained by binning the event-by-event Stokes parameters, calculated following \cite{Kislat2015}, according to the measured energy of the event. $Q$ and $U$ are given by the source flux multiplied by the polarization degree and either the cosine or the sine of twice the polarization angle, respectively. The Stokes $I$, $Q$ and $U$ energy spectra convolved with the instrument response are shown in the left, central and right panels, respectively. The best-fit model of the form \textsc{TBabs*(bbody * polconst + powerlaw * polconst)} is shown by the solid lines. Fit residuals are shown in the bottom panels; see Table~\ref{tab:spectropol_combined} in Appendix~\ref{appfit} for the fit parameters. Light colored curves are the background of IXPE detectors.
}
\label{fig:spectropol_powerlaw}
\end{figure*}

We tested two models commonly used to interpret magnetar soft X-ray spectra, i.e. an absorbed BB+PL and a BB+BB\footnote{Although BB radiation is not polarized, we assume that the shape of the thermal (polarized) spectra can be approximated by the Planck function.}. 
We started with the BB+PL model,  multiplying the BB and PL components with a polarization factor constant in energy; the corresponding XSPEC model is \textsc{TBabs*(bbody * polconst + powerlaw * polconst)}, with abundances from \citet{Wilms2000}. 
We include multiplicative constant factors for each \ixpe\ DU and for each observatory, except for \ixpe\ DU1 which is used as reference.
The fit gives a statistically acceptable $\chi^2=410.4$ for $408$~degrees of freedom (Figure~\ref{fig:spectropol_powerlaw}). The best-fit parameter values are reported in Table~\ref{tab:spectropol_combined} in Appendix~\ref{appfit}. Our fit parameters are in approximate agreement with those reported in \citet{rea+07}. The different neutral hydrogen column densities ($n_\mathrm H\sim 2 \times 10^{22}\ \mathrm{cm}^{-2}$ instead of $\sim 1.4\times 10^{22}\ \mathrm{cm}^{-2}$) largely result from the use of a different absorption model and elemental abundances, \textsc{TBabs} and \citet{Wilms2000} in our case instead of \textsc{phabs} and \citet{Anders1989}. We find however a significantly steeper PL  than  \citet{rea+07}. The PL  component dominates over the BB component at all energies, the latter contributing only $\sim20\%$ in the 2--8~keV energy range (see Fig.~\ref{fig:spectropol_powerlaw}). 

The simultaneous fit of the $Q$ and $U$ energy spectra provides further insights. We find that the PA associated to the BB component turns out to be different by $\sim 90^\circ$ with respect to that of the PL. This is similar to what \ixpe\ observed in 
\srcold{}  \citep[][]{taverna+22}, and it  supports a scenario in which thermal and non-thermal photons are polarized, one in the X and the other in the O mode. The changing polarization degree with energy can be then explained as due to the different relative contributions of the two orthogonally polarized components.
However, the polarization degree turns out to be much higher in  \srcshort{} than in \srcold. 
Taken at face value, the best fit gives a 100\% polarized blackbody component (see Table~ \ref{tab:spectropol_combined} in Appendix~\ref{appfit}). 
However, the polarization properties of this component are not well constrained and, for instance, fixing the polarization degree of the BB component to 20\% still gives
an acceptable fit ($\chi^2=434.2$ for $408$~degrees of freedom). 
We infer that the PL component is linearly polarized to $\sim$~65--75\%, depending on the polarization of the blackbody component. This polarization degree is much higher than predicted by the 
resonant scattering scenario \citep{taverna+22}. 
It should be noted that the assumptions of constant thermal 
and power law polarization degrees lead to a significant contribution of the 
thermal component. The data can be fit with a 
single PL component if we allow for a polarization degree with a linear
energy dependence (we get a $\chi^2=454.3$ for $410$~degrees of freedom for a power law fit with a photon index of $3.43 \pm 0.02$; 
the probability for getting higher $\chi^2$ values by chance 
is  $\sim 6\%$).

\begin{figure*}[th!]
\includegraphics[width=\textwidth]{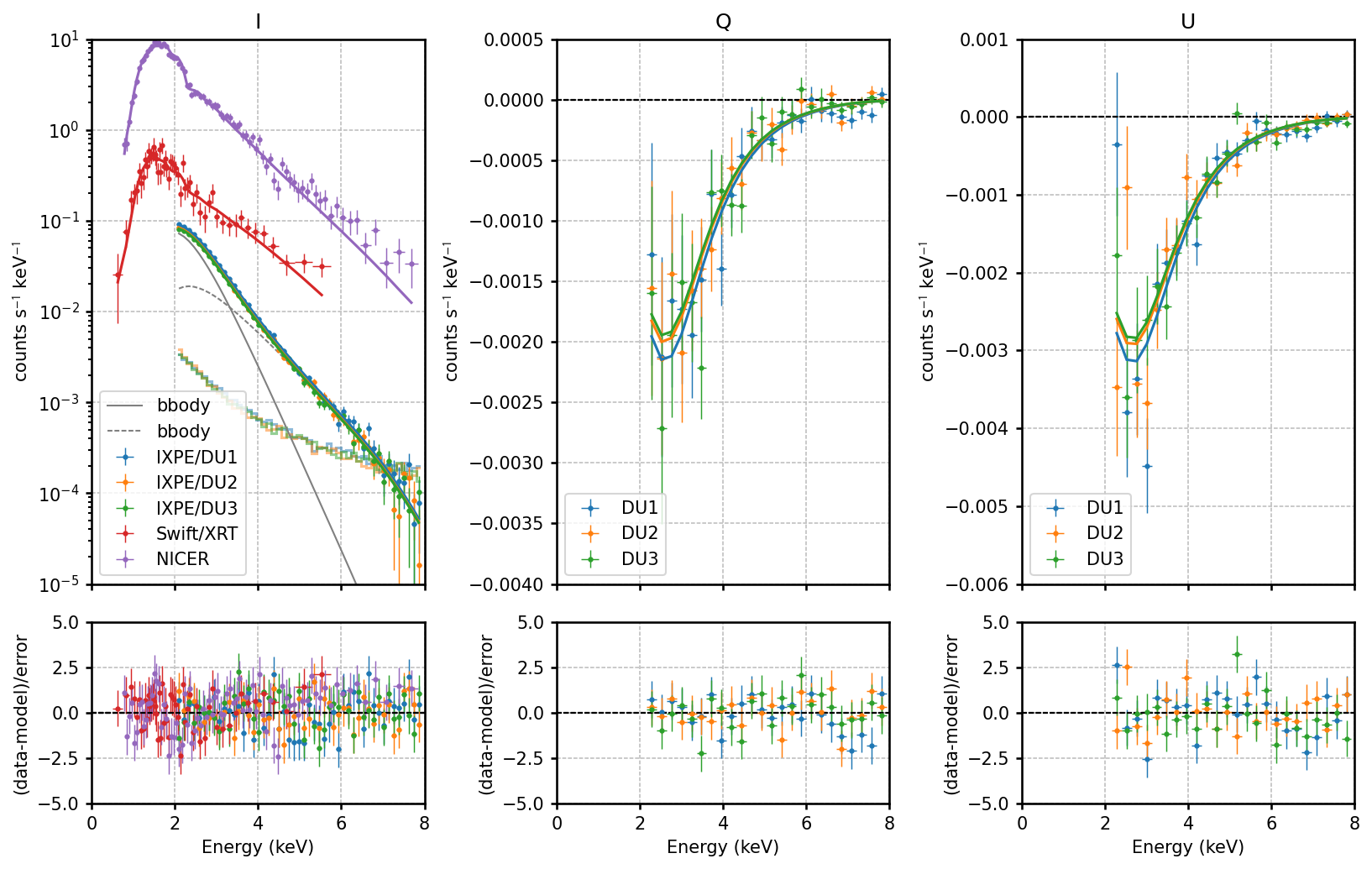} 
\caption{Same as in Figure~\ref{fig:spectropol_powerlaw}, for the  \textsc{TBabs*(bbody * polconst + bbody * polconst)} model. See Table~\ref{tab:spectropol_combined} in Appendix~\ref{appfit} for the fit parameters. \label{fig:spectropol_bbody}}
\end{figure*}

In the next step, we fitted the data with an absorbed BB+BB model, again assuming a constant polarization for each additive component; the correspondent XSPEC model is \textsc{TBabs*(bbody * polconst + bbody * polconst)}. The result of the fit is shown in Figure~\ref{fig:spectropol_bbody} (see again Appendix~\ref{appA},  Table~\ref{tab:spectropol_combined}). Also in this case, the fit is acceptable ($\chi^2=405.8$ for $408$~degrees of freedom) and the high-energy BB component is highly polarized, $\sim$~70\%. 
The low-energy component is polarized parallel to the high-energy component. The  best-fit model exhibits low polarization in the cold BB component, and by fixing the polarization degree to $\sim$~20\% also provides an acceptable fit
with $\chi^2=422.5$ for 409 degree of freedom. It is worth noting that  requiring a low-energy polarization angle orthogonal to the high-energy polarization angle gives
a low-energy polarization degree consistent with zero. Therefore, in this scenario, the observed increase of polarization with energy stems from the superposition of two parallel polarized components, a weakly polarized component dominating at low energies and a strongly polarized component dominating at high energies, in the \ixpe\ range.

\section{Phase-resolved spectropolarimetric analysis} 
\label{polphase}

\begin{figure*}[th!]
\includegraphics[width=\textwidth]{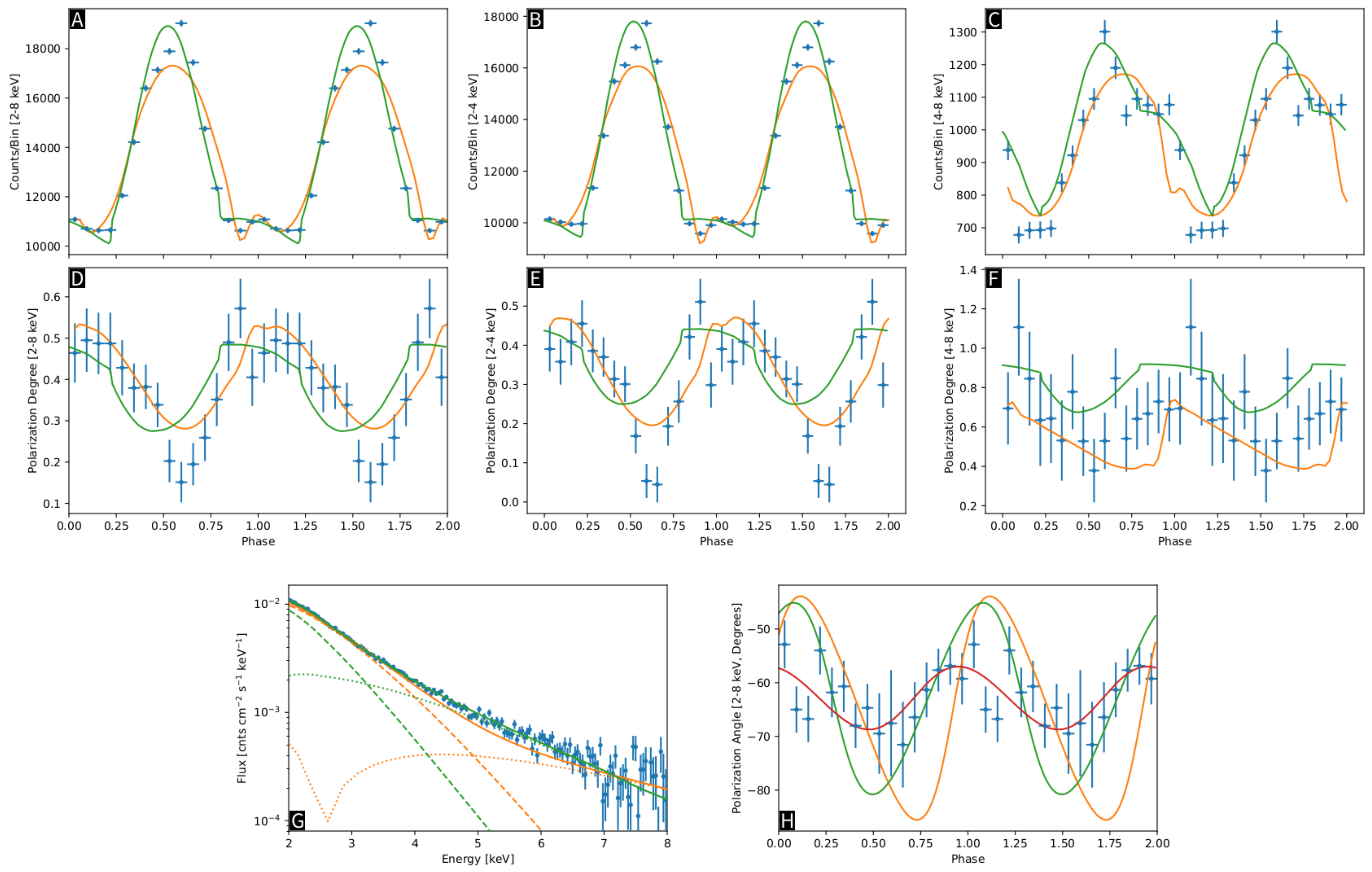}
\caption{
\srcshort\ \ixpe\ observation (blue circles with error bars) for the source counts  (panels A to C) and  polarization degree (panels D to F), plotted as a function of the rotational phase (in each row, the three panels from the left to the right show the data integrated over the $2$--$8$~keV, $2$--$4$~keV  and $4$--$8$~keV  bands, respectively). Panels G and H show:  (left)  the phase averaged spectrum (solid lines) together with the contributions of the condensed zone (dashed lines) and atmospheric zone (dotted lines);
(right) the polarization angle as a function of the rotational phase, integrated over the 2--8~keV band. Errors are shown at the 1 $\sigma$ level assuming the polarization degree and polarization angle are not correlated. In all panels the orange and green lines represent the model expectation within the belt+cap and cap+cap scenarios (solid lines for the total model, dotted and dashed lines in panel G only for the two separate components); the red solid line in panel H shows the RVM  (see the text for the parameter values and other details). Note that data points in the phase-resolved analysis are obtained with  \ixpeobssim, and therefore independent on the spectral decomposition. 
\label{fig:pol_phase}}
\end{figure*}

The first step of our phase-resolved analysis involved the determination of an accurate timing solution of the count rate data (see Appendix \ref{apptim}). 
Using epoch MJD 59850.84175 (TDB) as a reference, this provided
$f = 0.090795742(5)$\,Hz, $\dot{f}=-1.87(25)\times 10^{-13}$\,Hz s$^{-1}$.
The addition of the second frequency derivative did not improve the fit significantly.  
We then used this timing solution to perform the phase-resolved analysis. To this aim, we selected photons coming from different rotational phases, and repeated the procedures of Section~\ref{polspec} with the event files of the different phase bins. 
Rotational phases were ascribed to single events starting from their arrival times using the \ixpeobssim~tool \texttt{xpphase}, and events in the same phase bins were then  added together with the tool \texttt{xpselect}. Results are shown in Figure~\ref{fig:pol_phase} for the total ($2$--$8$ keV), low ($2$--$4$ keV) and high ($4$--$8$ keV) energy bands.
The source flux exhibits a single-peaked, nearly sinusoidal profile, as already reported by \cite{rea+03}; the pulse shape and the pulsed fraction change with energy. The polarization degree and angle also vary  with phase, and their pulse profiles are different in different energy bands. The pulse of the polarization degree is broadly anti-correlated with that of the flux  both in the total ($2$--$8$~keV) and low ($2$--$4$~keV) energy bands.

On the other hand, in the high energy interval ($4$--$8$~keV) the modulation of the signal is less significant and no robust conclusion can be reached whether this anti-correlation holds at these energies as well. 

\section{Discussion }\label{discuss}

\srcshort\ and the Vela PWN \cite[][]{xie+2022}
are the most strongly polarized \ixpe\ sources detected so far. 
The polarization degree of the phase-integrated emission of \srcshort\ increases from $\sim 20\%$ around $2$ keV to $\sim 80\%$ at $7$--$8$ keV. 
The polarization angle, on the other hand, stays roughly constant at $\sim 60^\circ$ (counted West of North). The polarimetric properties of \srcshort\ are quite at variance with those of the other magnetar observed by \ixpe.
For both sources, a BB+PL and a BB+BB model provide a satisfactory fit of the \ixpe\ spectrum (see \S\ref{polspec}). However, 
\srcold\ exhibits a lower polarization degree which decreases to approximately zero at $\sim 5$ keV, where the polarization angle swings by $90^\circ$, and then increases to $\sim 35\%$ around $8$ keV \cite[][]{taverna+22}.

The swing of polarization angle by $90^\circ$ and the presence of a power-law tail at higher energies with a value of polarization degree close to $30\%$ led \cite{taverna+22} to suggest that the X-ray emission from \srcold\ 
is in two different normal modes below and above $\sim 5$ keV. One possibility that has been suggested is that the low-energy photons are mostly polarized in the O mode and come from either a heated equatorial belt on the condensed neutron star surface or an atmospheric zone which is hit by back-flowing, bombarding magnetospheric currents \cite[][]{gonzalez2019},  while the high-energy ones are mainly in the X mode being reprocessed by magnetospheric RCS.

The picture for \srcshort\ is different. The very high degree of polarization observed at higher energies can not
be reproduced by the RCS mechanism, which indicates that RCS is not present, or, at least, that it does not affect primary photons much in the \ixpe\ energy range. 
The lack of a swing in the polarization angle points to the emission in the $2$--$8$ keV range  being dominated by a single mode. The relative contribution of the other mode is larger at lower energies, explaining the lower polarization degree at lower energies. If we assume that radiation comes from the magnetar surface, a polarization
degree as large as $\sim 70$--$80\%$ (as that observed at high energies) can be obtained if  the star is covered (at least in part) by an atmosphere \cite[e.g.][]{gonzalez2016, taverna+20,Caiazzo2022}. Emission, then, would be thermal and so the most likely spectral model, among those compatible with the data (see \S\ref{polspec}), is the BB+BB one.  
Standard atmospheric emission, however, is hard to reconcile with a polarization degree as low as $\sim 20\%$ (unless the viewing geometry is very particular), as
it is observed in \srcshort\ at low energies. It is nevertheless possible that, because of a non-homogeneous temperature distribution, some regions of the star surface are cold enough to undergo a phase transition that turns the atmosphere into a magnetic condensate. Emission from the condensed surface can be either in the X or in the O mode but the PD is always modest, $\lesssim 20\%$, and, if the condensate coexists with a hotter region that has an atmosphere on top, this could explain the \ixpe\ observation of \srcshort.

To explore this scenario, we ran a number of simulations assuming that the star surface comprises a hot atmospheric patch at temperature $T_{\rm hot}$ and a warm condensed region
at $T_{\rm warm}$; the rest of the surface is again condensed but at a much colder temperature $T_{\rm cold}$\footnote{This was done to ensure a physically consistent picture and in the actual modelling $T_{\rm cold}=0.15$ keV was assumed. However, the contribution of the colder zone does not affect the observed spectral and polarization properties.}. We tried several different geometries, leaving as free parameters the inclination  of the line-of-sight, $\chi$, and  of the magnetic axis, $\xi$, with respect to the spin axis. This, necessary oversimplified, picture is mostly meant to suggest possibilities on how the main feature of a complex thermal map may look like,  and not to provide a diagnostics of the shape and location of the emitting regions. 

The emission from the condensate is treated in the free-ion limit and for a Fe composition, following  \citet{potekhin+12}. The lack of observed spectral features in the soft X-rays argues against the presence of a heavy-elements atmosphere \cite[see the models by][for $B=10^{12}$--$10^{13}$~G]{Mori2007}. On the other hand, a light elements composition (mostly hydrogen) may be expected if the star experienced episodes of accretion/fallback. We computed emission from a pure hydrogen, completely ionized atmosphere using the numerical setup described in \citet{lloyd2003}\footnote{We caveat that this is an approximation, since strictly speaking these models can be applied only for $B\lesssim 10^{14}$~G.}. No mode conversion is assumed at the vacuum resonance \citep[see e.g.][]{Ho2003}.  
Photons were then propagated to the observer by using a ray-tracing method. The effects of vacuum birefringence as photons cross the star magnetosphere are accounted for.

We tested various configurations and found that there exist at least two geometrical set-ups, within the scenario outlined above, that match well most of the observed properties of \srcshort.
The first one (model A, or ``belt plus cap'') is a variation of the geometry envisaged for \srcold\ \cite[][]{taverna+22}, in which the warm (condensed) region corresponds to an equatorial belt, and the hot (atmospheric) region to a circular, polar cap; the belt, however, needs to be limited in azimuth in order to reproduce the X-ray pulse profile. 
In this case, calculations have been performed using the general relativistic ray-tracing code discussed in \citet[see also \citealt{taverna+15}]{zt06}. 
In the second model (model B, or ``cap plus cap'') the warm and hot regions are two circular spots, with the warm (condensed) one located  about $150^\circ$ from the spin axis, and a hot (atmospheric) spot located in the opposite hemisphere and displaced in longitude by about $90^\circ$ with respect to the warm one. For this latter scenario, we used the spectro-polarimetric magnetar model discussed in \citet{Caiazzo2022}, properly extended to account for different circular regions. We refer to the original papers for a detailed description of the models and the meaning of the various parameters. 

We found that the values of the parameters that best match the data are similar in the two cases:  $T_{\rm hot}\sim 0.8$ keV, $T_{\rm warm}\sim 0.6$ keV (all values are referred to the star surface), and a magnetic field of $\sim 10^{14}$~G (model B) or  $\sim 5 \times 10^{14}$~G (model A). 
Also the values of the geometrical viewing angles  are comparable, $\chi\sim 30^\circ$ and $\xi\sim 10^\circ$. The size of the emitting regions turns out to be somewhat smaller for model B, semi-aperture $\sim 5^\circ$ vs. $\sim 10^\circ$ for the atmospheric cap and semi-aperture $\sim 22^\circ$ of the warm cap as compared to $\sim 30^\circ\times 270^\circ$ of the belt. 
The observed pulsations require the size of the emitting regions to be relatively small, and this place an upper limit on  the source distance to  about 5~kpc. 
 
QED vacuum birefringence can strongly affect the polarization properties at infinity, leading  to an enhancement in the polarization degree of the observed signal.
However, the extent to which QED effects are measurable is actually quite sensitive to the size of the radiating region
on the star surface. As discussed in \cite{vanAdelsberg2009}, if emission comes from a small polar cap,
the depolarization due to geometrical effects is not present, so that expectations from models computed with or without vacuum birefringence are quite the same. In the case of \srcshort\, the small size of the two emitting regions (in particular that of the hot, highly polarized spot) does not allow for a robust test of vacuum birefringence. 

Both these models can reproduce the main spectropolarimetric features observed in \srcshort\ by \ixpe : namely the spectrum; the energy dependence of the phase-averaged polarization degree and angle across the $2$--$8$ keV band (see Fig.~\ref{fig:pol_vs_energy_new}); the pulse profile in the entire \ixpe \ range ($2$--$8$ keV), and in the $2$--$4$ and $4$--$8$ keV bands; the energy integrated polarization degree as a function of phase in the total, $2$--$4$ and $4$--$8$ keV ranges, including the anti-correlation in phase of the polarization degree and the flux in the total and low-energy bands (see Figure~\ref{fig:pol_vs_energy_new}). 
As it can be seen in the panel H of Figure \ref{fig:pol_phase}, the observed amount of swing in the
polarization angle with phase can also be explained by both models; the red curve depicts the
best-fit
simple rotating vector model \citep[RVM; with unbinned likelihood
analysis, e.g.][]{gonzalez2022}. 
In both the proposed configurations model spectra show no features, in agreement with data (see Figure~\ref{fig:pol_phase}). In principle absorption features are expected in the spectrum of  a pure hydrogen atmosphere or in that of the magnetic condensate. However, the strong proton cyclotron line in the atmospheric component is  $\sim0.6$~keV in model B, and is drowned out by the dominant contribution from the colder belt in model A.  Emission from the condensate component predicts spectral features at low energies ($\sim 0.6$~keV) but, at least for these model parameters, they are too faint to be detectable in the currently available spectra.

The observed  anti-correlation in phase of the polarization degree and the flux is due to the relative dominance of two regions, and to the fact that the (less polarized) condensed zone is contributing to much of the counts below $\sim 4$ keV. 

There are, however, some caveats. The surface temperature of the warm condensed region is very close to that of the hotter atmospheric patch. Although the conditions for a phase transition
in a strongly magnetized medium are still largely uncertain,  assuming a dipolar magnetic field with polar strength $\sim5 \times 10^{14}$~G,
condensation is expected at $T\lesssim 0.5$ and $T\lesssim 0.9$ keV at the equator and the pole, respectively,
for a Fe composition \cite[][see also figure 1 in \citealt{taverna+20}]{medin+lai07}. This may
indicate that, if the  hot and warm regions at the star surface, which are in different phases, have the same chemical composition, they are very close to the phase transition. Alternatively, it may indicate that  the chemical composition varies across the surface (as in our models, which assume a pure hydrogen atmosphere and an iron condensate),  that the  
real surface temperature map is more complicated than what is caught in our, necessarily, oversimplified model, or that  there may be  departures of the local magnetic field from the (overall) dipole topology.

As mentioned before, the spectro-polarimetric analysis indicates that the data are also compatible with a scenario in which emission is dominated by a non-thermal (PL) component. Would that be the case, the fact that the polarization degree decreases at low energies may be due either to the presence of a second sub-dominant spectral component with emission in the opposite polarization mode (e.g. a thermal emission from a condensed part of the crust) or to the fact that the non-thermal emission has an intrinsic variation of the polarization properties with energy. The  main reason why we have not attempted to explore this scenario is that the very high degree of polarization observed at high energy ($\sim 70-80\%$ ) cannot be reconciled with the physical models commonly proposed to explain the non-thermal emission from magnetars in the soft X-ray band, e.g. RCS (saturated RCS  only predicts a $\sim30\%$ level of polarization). However, it is worth noticing that \srcshort\ exhibits a strong non-thermal emission at high energies, as detected in the {\it INTEGRAL} band,  with  extreme variations in phase and energies, whose origin is still unexplained in terms of detailed physical models \cite[][]{kuiper+1996}.
It could therefore be possible that this non thermal component and that observed in the \ixpe\ band could be similar in origin (but with varying degrees of manifestation).

\begin{acknowledgments}
The Imaging X-ray Polarimetry Explorer (IXPE) is a joint US and Italian mission.  The US contribution is supported by the National Aeronautics and Space Administration (NASA) and led and managed by its Marshall Space Flight Center (MSFC), with industry partner Ball Aerospace (contract NNM15AA18C).  The Italian contribution is supported by the Italian Space Agency (Agenzia Spaziale Italiana, ASI) through contract ASI-OHBI-2017-12-I.0, agreements ASI-INAF-2017-12-H0 and ASI-INFN-2017.13-H0, and its Space Science Data Center (SSDC) with agreements ASI-INAF-2022-14-HH.0 and ASI-INFN 2021-43-HH.0, and by the Istituto Nazionale di Astrofisica (INAF) and the Istituto Nazionale di Fisica Nucleare (INFN) in Italy.  This research used data products provided by the IXPE Team (MSFC, SSDC, INAF, and INFN).  
We thank K. C. Gendreau and Z. Arzoumanian for their help in scheduling NICER observations of the source. We acknowledge the use of public data from the Swift data archive, and we thank the Swift team for promptly scheduling a ToO observation.
R. Ta. and R. Tu. acknowledge financial support from the Italian MUR through grant PRIN 2017LJ39LM.
D.G.-C., J.H. and I.C. acknowledge support from the Natural Sciences and Engineer Council of Canada and the Canadian Space Agency.
M.Negro acknowledges the support by NASA under award number 80GSFC21M0002.T.T. was supported by grant JSPS KAKENHI JP19H05609. HK and EG acknowledge NASA support under grants 80NSSC18K0264, 80NSSC22K1291, 80NSSC21K1817, and NNX16AC42G. We thank an anonymous referee for a careful reading of the paper.

\end{acknowledgments}

\facilities{\ixpe, Swift(XRT), NICER}





\appendix
\section{Observational data and data processing}
\label{appA}
\subsection{IXPE} \label{ixpedata} 

\ixpe\  observed \srcshort\ in two segments close in time, from 2022-09-19 05:08 UTC to 2022-09-29 12:00 UTC and from 2022-09-30 12:52:02 UTC to 2022-10-08 11:17, for a total livetime of $\sim 837$~ks. Level 2 (LV2) data, processed to produce suitable inputs for science analysis, were downloaded from the \ixpe\ archive at HEASARC\footnote{https://heasarc.gsfc.nasa.gov/docs/ixpe/archive/}. One file for each of the three \ixpe\ telescopes was generated, and they were further processed independently. Photons arrival times are corrected to the solar system barycenter in the Barycentric Dynamical Time (TDB) scale, with the \textsc{FTOOL/barycorr} tool, included in HEASOFT 6.31, using the object coordinates stored in the LV2 file, the Jet Propulsion Laboratory Development Ephemeris 421 and the International Celestial Reference System reference frame. 
Background events in the LV2 files are at first partially rejected. They are identified starting from topological characteristics of the event, e.g., the number of hit pixels and the fraction of the energy in the main track with respect to the total collected for the event. Events falling outside pre-defined intervals are tagged as background and removed. This approach  allows to reduce the background counting rate by more then $30\%$, at the cost of a reduction of the source count rate by $\sim 0.5\%$, which is deemed negligible for the subsequent analysis \cite[see][for a more detailed description]{DiMarco2023}.

\begin{figure*}[th!]
\includegraphics[width=\textwidth]{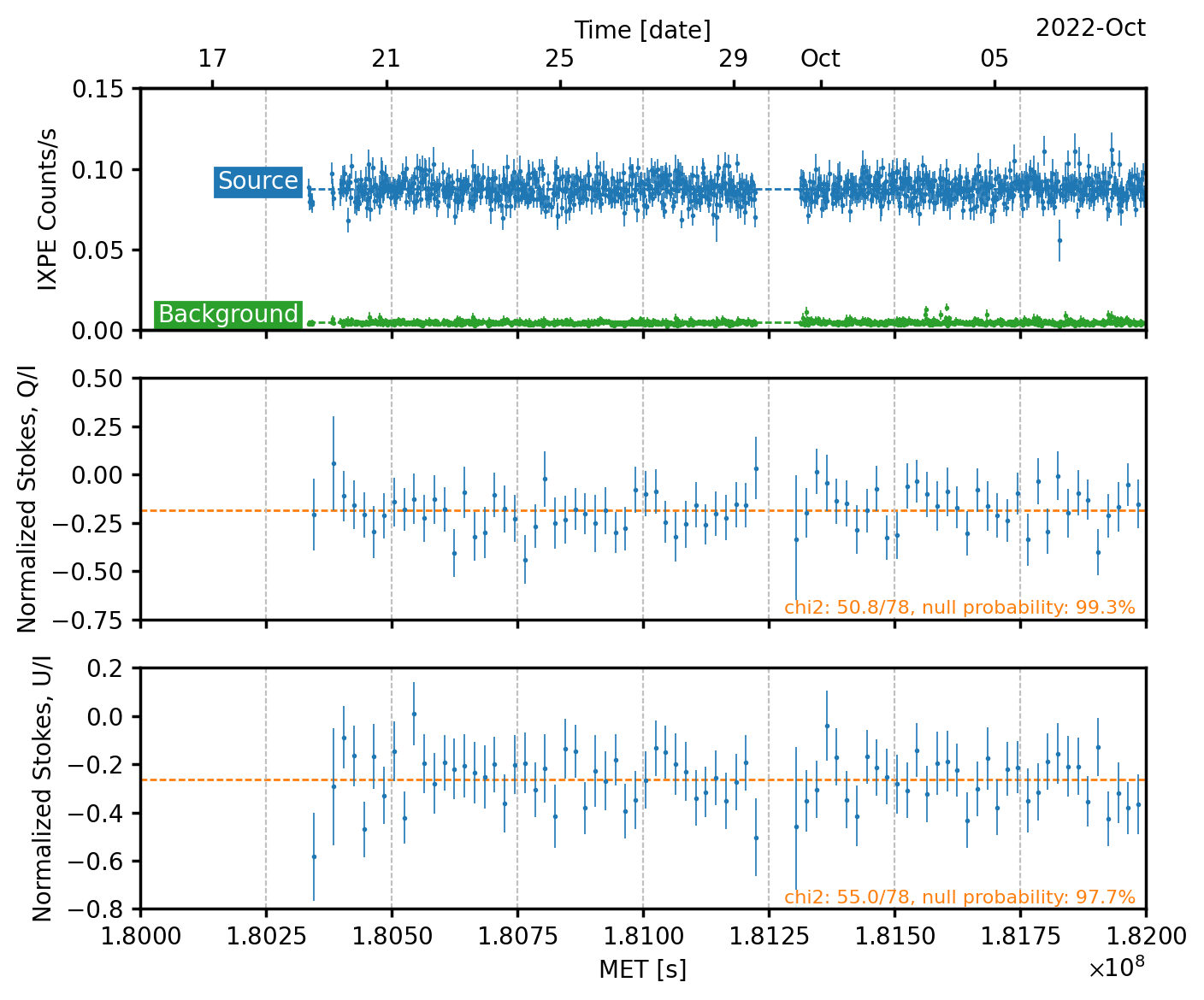} 
\caption{ Top panel: \srcshort\ light curve, averaged over the three \ixpe\ telescopes, for the source and the background. The count rate is averaged over 500~seconds and integrated over the \ixpe\ nominal energy band, $2.0$--$8.0$ ~keV. Middle and bottom panels: light curve of normalized Stokes parameters; the time bin is 20 ks. In all panels variations are compatible with statistical fluctuations only: the null probabilty for a fit with a constant is acceptable.
\label{fig:lightcurve}}
\end{figure*}

The source region is then defined with SAOImage DS9 \citep{Joye2003} as a circle with radius of 1.5~arcmin. The background is extracted from an annulus centered on \srcshort\ with inner and outer radii of 2.5 and 4.0~arcmin, respectively. Response to polarization is extracted from LV2 data with the \ixpeobssim\  package \citep{Baldini2022}, which 
implements the algorithms in \citet{Kislat2015} and is publicly available\footnote{https://github.com/lucabaldini/ixpeobssim}. 
Spectro-polarimetric analysis is carried out by feeding the Stokes spectra, generated with the \ixpeobssim\ tool \textsc{xpbin}, into \textsc{XSPEC} \citep{Arnaud1996} and exploiting the appropriate response matrices available at HEASARC for modelling the polarization as suggested by \citet{Strohmayer2017}. The light curve of \src{} and of the background during the two \ixpe\ observations is shown in Figure~\ref{fig:lightcurve}. Note that the first segment is splitted in two pointings, the first between 19-09-2022 5.00 - 7:20 UTC and the second starting at 19-09-2022 17:30 and ending at 29-09-2022 12:00. Both components did not show any indication of either a spectral or polarization variation with time. To test this, we binned the lightcurves of the Stokes parameters and background in bins of 20ks and we checked that the variations are compatible with statistical fluctuations only (the null probabilty for a fit with a constant is acceptable). Therefore, we proceeded with the analysis by summing all the data.

\subsection{NICER} \label{sec:nicerdata}

In order to complement the \ixpe\  data, we requested a set of new NICER observations, which have been awarded and performed a week after the end of the last \ixpe\  observation. However, all these  observations suffered from high optical loading (resulting from a low Sun angle), and thus were unusable for our analysis. We therefore resorted to the use of the NICER archival data, using the most recent available observations (closest in time to the \ixpe\  observation). Specifically, for the contemporaneous spectral fitting with IXPE and Swift spectra, we made use of the observations from 2022-08-21 (ObsID 5593051301).

The data were processed using version 9 of the NICER Data Analysis Software (\texttt{NICERDAS}) on version 6.30.1 of \texttt{HEASoft}.  We applied standard filtering criteria as per the default cuts with \texttt{nicerl2}, resulting in 972 s of filtered exposure. The spectrum was then binned with optimal binning \citep{Kaastra16} and with a minimum of 25 counts per bin. The background spectrum was generated using version 7 of the 3C50 model \citep{Remillard22}. The rmf and arf files were generated from \texttt{nicerrmf} and \texttt{nicerarf}, respectively.

\subsection{Swift} \label{sec:swiftdata}

A Swift-XRT target of opportunity observation was performed on  2022-10-20 for a total of 1020\,s in Windowed Timing (WT) mode. The data were processed using the standard HEASoft tools, extracting the spectra using \texttt{xselect}, after generating ancillary response file using \texttt{xrtmkarf} and the latest calibration files available in the Swift-XRT CALDB. The source was extracted from the cleaned event file using a centered circular region with a radius of 17 pixels. The background region was extracted using an off-centered circular region of the same radius. The extracted source spectra were binned, requiring each spectral channel to have at least 25 counts.

\section{Model parameters from the spectro-polarimetric analysis}
\label{appfit}

As discussed in \S~\ref{polspec}, there are two possible spectral decompositions that are both compatible with the data: one consists of a thermal (blackbody) component plus a non thermal (power law) and the second consists of two blackbodies. The best fitting model parameters, for the two cases, are listed in Table \ref{tab:spectropol_combined}.

\begin{table}
\begin{tabular}{c|c|| c | c }
Model parameters & value  &  model parameters  & 
 value \\
\hline
 DU1 normalization  & 1.0 (frozen) &
 DU1 normalization  & 1.0 (frozen) \\ 
 TBabs nH [ 10$^{22}$ cm$^{-2}$] & 2.085$^{+0.055}_{-0.050}$   & 
 TBabs nH [ 10$^{22}$ cm$^{-2}$] & 1.391$^{+0.039}_{-0.038}$  \\
 kT$_\textrm{bbody}$ [ keV ] & 0.4546$^{+0.0077}_{-0.0058}$  & 
  kT$_\textrm{bbody1}$ [ keV ] & 0.4354$^{+0.0076}_{-0.0078}$  \\
 norm$_\textrm{bbody}$  & 0.0002293$^{+0.0000179}_{-0.0000096}$  & 
 norm$_\textrm{bbody1}$  & 0.0005440$^{+0.0000097}_{-0.0000091}$  \\
 PD$_\textrm{bbody}$  & 1.0  (U.L.) $^\ddag$  & PD$_\textrm{bbody1}$  & 0.060$^{+0.036}_{-0.037}$  \\ 
PA$_\textrm{bbody}$ [ deg ] & 30.5$^{+3.4}_{-3.3}$  & 
PA$_\textrm{bbody1}$ [ deg ] & $-69.0^{+16.7}_{-18.2}$  \\ 
 PowLaw Photon Index  & 2.9672$^{+0.0092}_{-0.0567}$  & 
 kT$_\textrm{bbody2}$ [ keV ] & 1.073$^{+0.031}_{-0.029}$  \\ 
 norm$_\textrm{PowLaw}$  & 0.04101$^{+0.00059}_{-0.00382}$  & 
 norm$_\textrm{bbody2}$  & 0.0002324$^{+0.0000082}_{-0.0000079}$  \\ 
 PD$_\textrm{power law}$  & 0.768$^{+0.041}_{-0.040}$  & 
 PD$_\textrm{bbody2}$  & 0.728$^{+0.038}_{-0.037}$  \\ 
 PA$_\textrm{power law}$ [ deg ] & $-60.8^{+1.0}_{-1.2}$  &
 PA$_\textrm{bbody2}$ [ deg ] & $-61.0^{+1.4}_{-1.4}$  \\ 
 DU2 normalization  & 0.9567$^{+0.0051}_{-0.0049}$  & DU2 normalization  & 0.9568$^{+0.0053}_{-0.0052}$  \\ 
 DU3 normalization  & 0.8921$^{+0.0048}_{-0.0046}$  & 
 DU3 normalization  & 0.8922$^{+0.0049}_{-0.0049}$  \\ 
 Swift/XRT normalization  & 0.922$^{+0.032}_{-0.032}$  & 
 Swift/XRT normalization  & 0.934$^{+0.032}_{-0.032}$  \\
 NICER normalization  & 1.062$^{+0.013}_{-0.011}$  & 
 NICER normalization  & 1.075$^{+0.013}_{-0.013}$  \\ 
 $\chi^2$ & 410.4 with 408 d.o.f.   & 
 $\chi^2$ & 405.8 with 408 d.o.f. \\
 Null prob & 45.7\% & 
 Null prob & 52.2\% \\
 \end{tabular}
\caption{Model parameters for the joint spectropolarimetric fit of \ixpe\, Swift/XRT and NICER data. Uncertainties are calculated at the 68.3\% confidence level for one parameter of interest. The two leftmost  columns refer to the model \textsc{TBabs*(bbody * polconst + powerlaw * polconst)}. The normalization of the BB components is defined as $L_{39}/D_{10}^2$, where $L_{39}$ is the BB luminosity in units of
$10^{39}$ erg/s and $D_{10}$ is the distance to the source in units of
10~kpc.
The PL normalization is in units of counts~keV$^{-1}$~cm$^{-2}$s$^{-1}$ at 1 keV.
The rightmost two columns are for the model \textsc{TBabs*(bbody * polconst + bbody * polconst)}. 
\newline $^\ddag$  In this case, the confidence level for the polarization degree of the BB component is not closed, and it was only possible to set an upper limit (U.L.). The value of this parameter is poorly constrained: by freezing the value of PD$_{\textrm{bbody1}}$ to 0.2 also returns an acceptable fit (see the text for details).}
\label{tab:spectropol_combined}
\end{table}

\section{Timing Solution}
\label{apptim}

A timing solution of the count rate data, has been obtained by  following  the procedure from \citet{bachetti2022}. We first determined  an approximate solution and then used the \citet{Pletsch2015}  Maximum Likelihood method to refine it.
We determined an approximate first solution with the tool \texttt{HENzsearch}, running a $Z^2_3$ search \citep{Buccheri1983} around the known pulse frequency from \citet{DibKaspi2014}.
The pulsar was detected with extremely high significance, with $Z^2_3\sim16000$ ($\sim$~126$\sigma$, considering 10000 trials), at $f=0.09079575$ Hz and $\dot{f}=-2.28\times 10^{-13}$\,Hz s$^{-1}$.
Then, we used the tool \texttt{ell1fit}\footnote{\url{https://github.com/matteobachetti/ell1fit}} v. 0.2 to run a Maximum Likelihood fit of the spin solution, using a template for the pulse profile obtained by a Fourier analysis with 4 harmonics, and letting only the spin frequency and its first spin derivative vary. 
The resulting solution, using epoch MJD 59850.84175 (TDB) as a reference, is 
$f = 0.090795742(5)$\,Hz, $\dot{f}=-1.87(25)\times 10^{-13}$\,Hz s$^{-1}$.
The addition of the second frequency derivative did not improve the fit significantly: the 3 $\sigma$ upper limit on the second derivative in our observation is $\sim2\times 10^{-19}$\,Hz s$^{-2}$. 

\bibliography{sample631}{}
\bibliographystyle{aasjournal}
\end{document}